# Multiple equilibrium states of a curved-sided hexagram: Part II—Transitions between states[1]

Lu Lu[a], Jize Dai[a], Sophie Leanza[a], John W. Hutchinson[b], Ruike Renee Zhao[a,*]

[a] Department of Mechanical Engineering, Stanford University, Stanford, CA 94305, USA
[b] School of Engineering and Applied Sciences, Harvard University, Cambridge, MA 02138, USA

*Corresponding author. Email: rrzhao@stanford.edu

**Abstract**

Curved-sided hexagrams with multiple equilibrium states have great potential in engineering applications such as foldable architectures, deployable aerospace structures, and shape-morphing soft robots. In Part I, the classical stability criterion based on energy variation was used to study the elastic stability of the curved-sided hexagram and identify the natural curvature range for stability of each state for circular and rectangular rod cross-sections. Here, we combine a multi-segment Kirchhoff rod model, finite element simulations, and experiments to investigate the transitions between four basic equilibrium states of the curved-sided hexagram. The four equilibrium states, namely the star hexagram, the daisy hexagram, the 3-loop line, and the 3-loop "8", carry uniform bending moments in their initial states, and the magnitudes of these moments depend on the natural curvatures and their initial curvatures. Transitions between these equilibrium states are triggered by applying bending loads at their corners or edges. It is found that transitions between the stable equilibrium states of the curved-sided hexagram are influenced by both the natural curvature and the loading position. Within a specific natural curvature range, the star hexagram, the daisy hexagram, and the 3-loop "8" can transform among one another by bending at different positions. Based on these findings, we identify the natural curvature range and loading conditions to achieve transition among these three equilibrium states plus a folded 3-loop line state for one specific ring having a rectangular cross-section with a height to thickness ratio of 4. The results obtained in this part also validate the elastic stability range of the four equilibrium states of the curved-sided hexagram in Part I. We envision that the present work could provide a new perspective for the design of multi-functional deployable and foldable structures.

---

[1] This paper is dedicated to Prof. Huajian Gao in recognition of his significant contributions to the solid mechanics community and to celebrate his 60th birthday. Ruike Renee Zhao deeply appreciates the invaluable mentorship and support provided by Prof. Gao.

## 1. Introduction

Slender structures are widely observed in nature and are regularly employed in engineering applications across various length scales. Pertinent examples include curled hairs (Audoly and Pomeau, 2010; Miller et al., 2014), overhand knots (Jawed et al., 2015), and medical guidewires (Wang et al., 2021). Among various slender structures, ring origami (Lu et al., 2023b; Wu et al., 2021a), formed by closed-loop rods of different geometries (e.g., circular rings (Audoly and Seffen, 2015; Box et al., 2020; Manning and Hoffman, 2001; Yoshiaki et al., 1992) and polygonal rings (Wu et al., 2022; Wu et al., 2021a)), has recently emerged as a powerful platform to design foldable and deployable structures with high packing abilities, such as foldable solar panel devices (Leanza et al., 2022), foldable tents (Mouthuy et al., 2012), and deployable space structures (Pellegrino, 2001). By harnessing the snap-through instability, circular or polygonal rings with rationally designed geometric parameters can fold into a multi-loop overlapping state in a self-guided manner under sufficient external bending, twisting, or point loads (Lu et al., 2023a; Sun et al., 2022; Yoshiaki et al., 1992). However, it has been shown that these circular or polygonal rings can only exhibit monostable or bistable behavior (Sun et al., 2022; Wu et al., 2021a).

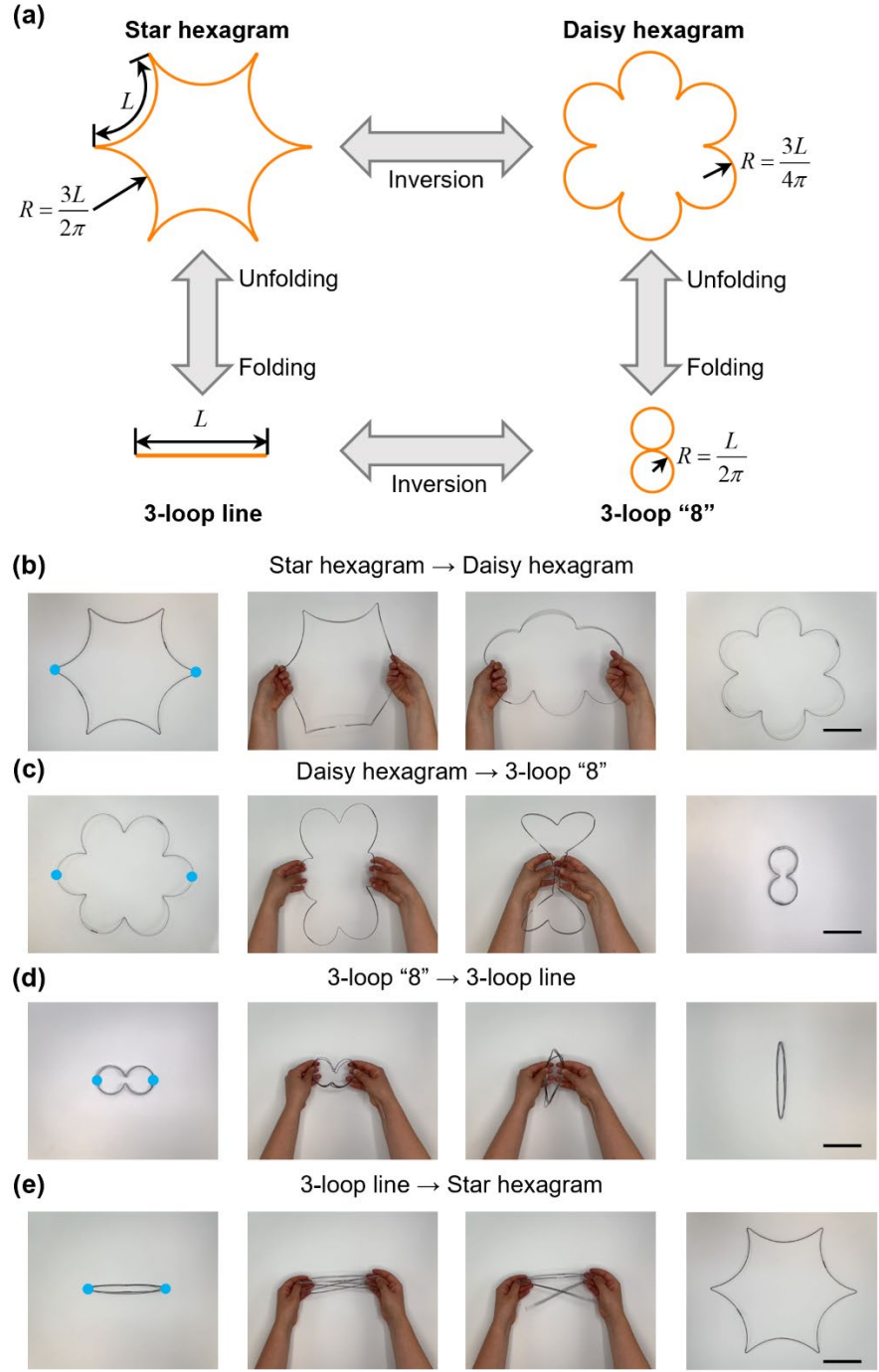

**Fig. 1.** Transitions between the four equilibrium states of the curved-sided hexagram made of stainless-steel rods having a rectangular cross-section with a height to thickness ratio of 4 and a natural curvature discussed in the paper. (a) Schematics and transition paths between the four equilibrium states. (b) Inversion from the star hexagram to the daisy hexagram. (c) Folding from the daisy hexagram to the 3-loop "8". (d) Inversion from the 3-loop "8" to the 3-loop line. (e) Unfolding from the 3-loop line to the star hexagram. The blue dots denote the loading positions. Scale bars: 10 cm.

Recently, Dai and Leanza et al. discovered that a curved-sided hexagram with a certain initial curvature can 'magically' fold into a 3-loop line (Dai et al., 2023; Leanza et al., 2023). More interestingly, here we have found that by rationally designing its natural curvature (the curvature in its disconnected state), the curved-sided hexagram exhibits more equilibrium states. The four basic equilibrium states of the curved-sided hexagram are illustrated in Fig. 1(a), which are referred to as the star hexagram, the daisy hexagram, the 3-loop line, and the 3-loop "8". The rods comprising the edges of the rings are treated as being inextensional and of length $L$. In particular, the star hexagram consists of six 120° circular arcs of radius $R=3L/2\pi$, and the daisy hexagram is composed of six 240° circular arcs of radius $R=3L/4\pi$. These two equilibrium states can transition between each other through inversion, namely one state's inner surface is turned into the outer surface of the other state. Also, they can transform into the other two compact states through folding. The star hexagram can fold into a 3-loop line which has an idealized area of zero, with length $L$. The daisy hexagram can fold into a 3-loop "8", with both circles in the "8" having length $L$ and radius $L/2\pi$. Also, the 3-loop line can transform into the 3-loop "8" through inversion and vice versa. Transitions between the four equilibrium states can be achieved by applying appropriate external stimuli. For example, starting with the inversion of the star hexagram into the daisy hexagram by bending at two corners (Fig. 1(b)), the daisy hexagram can then be folded into a 3-loop "8" by bending at two edges (Fig. 1(c)). Subsequently, the 3-loop "8" can be inverted into the 3-loop line by bending at its edges (Fig. 1(d)). Finally, the 3-loop line can be unfolded back into the star hexagram (Fig. 1(e)). An experimental demonstration of this example is presented in Video 1 in the Supplementary Materials, and experimental details on how to fabricate these rings are provided in Fig. S1 in the Supplementary Materials.

Elastic structures that have multiple stable equilibrium states (i.e., multistable structures) have widespread engineering applications ranging from shape-reconfigurable architectures (Melancon et al., 2021), shape-morphing soft robots (Chi et al., 2022; Wu et al., 2021b), energy trapping devices (Shan et al., 2015), mechanical metamaterials (Liu et al., 2023b; Meng et al., 2022) to flexible electronics (Fu et al., 2018; Zhang et al., 2021). Benefiting from the multiple equilibrium states, these functional structures can switch from one functional state to another when subjected to a stimulus that can overcome the energy barrier between two states. Understanding the elastic stability and transitions between different states of multistable structures is important to guide the design and realization of multifunctional applications. In recent years, extensive efforts have been devoted to studying the mechanical behavior of multistable structures or systems such as multistable metamaterials formed from curved beams (Che et al., 2017) or curved shells (Liu et al., 2023a; Vasios et al., 2021), multistable corrugated shells (Norman et al., 2008) and foldable cones (Seffen, 2021), and multistable origami (Feng et al., 2020; Li and Pellegrino, 2020; Novelino et al., 2020; Waitukaitis et al., 2015). However, far less attention has been paid to the elastic stability and transitions of multistable systems formed by closed-loop slender rods (Yu et al., 2021).

In the companion paper, Part I, we have studied the elastic stability of the multiple equilibrium states of the curved-sided hexagram and obtained the natural curvature range within which each of the states is stable for rods with circular or rectangular cross-sections. In this paper, we will investigate the transitions between different equilibrium states of the curved-sided hexagram based on a combination of a multi-segment Kirchhoff rod model, finite element analysis (FEA), and experiments. For each equilibrium state (i.e., the star hexagram, the daisy hexagram, the 3-loop line, and the 3-loop 8), we comprehensively examine its transition behavior under external bending loads at corners or edges within the stability range, and present the transition processes, transition states, as well as moment and energy curves to understand the effects of natural curvature and loading position on the transitions of the four equilibrium states. Finally, to bring these ideas together, we provide strategies to achieve the transitions between the four equilibrium states for one particular curved-sided hexagram. The multiple equilibrium states of the curved-sided hexagram studied in Parts I and II provide a new perspective for the design of multistable, foldable, and reconfigurable structures, which have great potential in aerospace, robotics, and other engineering applications.

The remainder of this paper is organized as follows. In Section 2, we develop a unified multi-rod system based on the Kirchhoff rod theory, which is used to analyze the transition behavior of the four equilibrium states of the curved-sided hexagram under external bending loads. In Section 3, we study the transition behavior of each of the four equilibrium states for a range of natural curvatures under prescribed rotations at selected corners or edges. In Section 4, we propose strategies to achieve the transitions among the four stable equilibrium states of a specific curved-sided hexagram. In Section 5, we summarize the main contributions of this work.

## 2. Theoretical model

In this section, we present a multi-segment rod model based on the Kirchhoff rod theory to study the transition behavior of the four equilibrium states of the curved-sided hexagram under external stimuli. The schematic of an elastic rod of length $L$ is shown in Fig. 2(a). The rod is assumed to be inextensible and unshearable. In other words, the length of the rod remains unchanged, and the centerline of the rod remains perpendicular to the cross-section during deformation. To capture the deformation of the centerline, a position vector $\mathbf{p}(s)=p_1\mathbf{E}_1+p_2\mathbf{E}_2+p_3\mathbf{E}_3$ defined in the global material frame ($\mathbf{E}_1$, $\mathbf{E}_2$, $\mathbf{E}_3$) is introduced, in which $s$ is the arc length coordinate. A local material frame ($\mathbf{e}_1$, $\mathbf{e}_2$, $\mathbf{e}_3$) is attached to the centerline of the rod, in which $\mathbf{e}_1$ and $\mathbf{e}_2$ are unit vectors in the height and thickness directions, respectively, and $\mathbf{e}_3$ is the unit tangent vector of the centerline. The reader should be aware that some of the conventions and notation of Part I differ from those of Part II due to the fact that this two-part paper brings together two distinct approaches in the literature. For consistency with prior work, Part I adopts the same notation as its predecessor, Leanza et al. (2023), while Part II follows its predecessors, Sun et al., (2022) and Lu et al., (2023). In particular, in Part I, the tangent to the rod is $\mathbf{e}_2$ while in Part II it is $\mathbf{e}_3$. These differences will only appear in Section 2 and the Appendix of Part II. In the other sections of Part II, where the results will be presented and discussed, the notation will be the same as that in Part I. In particular, the sign of the natural curvature, $\kappa_n$, in Sections 3–5 of Part II follows the convention used in Part I. The kinematics of the local material frame is given by $\mathbf{e}'_i = \boldsymbol{\omega} \times \mathbf{e}_i$, and $\boldsymbol{\omega}=\kappa_1\mathbf{e}_1+\kappa_2\mathbf{e}_2+\kappa_3\mathbf{e}_3$ is the Darboux vector, with $\kappa_1$, $\kappa_2$, and $\kappa_3$ denoting the two bending curvatures and the twisting curvature during deformation, respectively. Note that $(\bullet)' = d(\bullet)/ds$ throughout the present paper. The matrix form of the kinematic equations can be written as

$$\begin{bmatrix} \mathbf{e}'_1 \\ \mathbf{e}'_2 \\ \mathbf{e}'_3 \end{bmatrix} = \begin{bmatrix} 0 & \kappa_3 & -\kappa_2 \\ -\kappa_3 & 0 & \kappa_1 \\ \kappa_2 & -\kappa_1 & 0 \end{bmatrix} \begin{bmatrix} \mathbf{e}_1 \\ \mathbf{e}_2 \\ \mathbf{e}_3 \end{bmatrix} = [\mathbf{K}] \begin{bmatrix} \mathbf{e}_1 \\ \mathbf{e}_2 \\ \mathbf{e}_3 \end{bmatrix}. \tag{1}$$

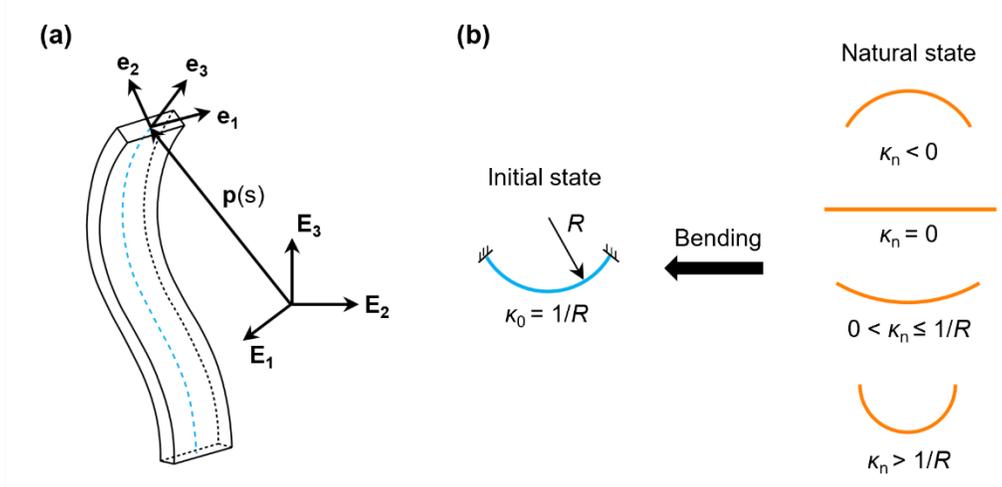

**Fig. 2.** Schematics of the rod model. (a) The local material frame ($\mathbf{e}_1$, $\mathbf{e}_2$, $\mathbf{e}_3$) is located on the centerline of the rod, and the position vector $\mathbf{p}(s)$ is defined in the global material frame ($\mathbf{E}_1$, $\mathbf{E}_2$, $\mathbf{E}_3$). (b) Natural state and initial state of an elastic rod. The rod with natural curvature $\kappa_n$ is bent into an arc having a uniform radius of curvature, $\kappa_0 = 1/R$, and is clamped at the two ends.

Here, we consider the rod has a uniform radius of curvature, $\kappa_0 = 1/R$, about $\mathbf{e}_1$ in the initial state, and a uniform curvature $\kappa_n$ about $\mathbf{e}_1$ in the natural state, as shown in Fig. 2(b). The natural curvature can be greater than, less than, or equal to the initial curvature. For a rod with a straight natural state, $\kappa_n = 0$. For a curved rod with its bending direction in the natural state opposite to the bending direction in the initial state, $\kappa_n < 0$. When the initial curvature is different from the natural curvature, the rod carries a uniform bending moment $M_n$ in its initial state, which equals the moment required to bend the rod from its natural state to the initial state, namely $M_n = EI_1(\kappa_0 - \kappa_n)$, with $E$ being the Young's modulus and $I_1$ being the moment of inertia of the cross-section with respect to $\mathbf{e}_1$. According to the Kirchhoff rod theory, in the absence of body forces and external couples, the balance of internal forces $\mathbf{N}$ and moments $\mathbf{M}$ gives

$$\mathbf{N}' = \mathbf{0}, \quad \mathbf{M}' + \mathbf{e}_3 \times \mathbf{N} = \mathbf{0}, \tag{2}$$

where $\mathbf{N} = N_1\mathbf{e}_1 + N_2\mathbf{e}_2 + N_3\mathbf{e}_3$ and $\mathbf{M} = (M_1 + M_n)\mathbf{e}_1 + M_2\mathbf{e}_2 + M_3\mathbf{e}_3$. For the linear constitutive relations used in this paper, $M_1 = EI_1(\kappa_1 - \kappa_0)$, $M_2 = EI_2\kappa_2$, and $M_3 = GJ\kappa_3$, where $G$ is the shear modulus, $I_2$ is the cross-sectional moment of inertia with respect to $\mathbf{e}_2$, and $J$ is the torsional constant. For a rod with rectangular cross-section, the two bending rigidities and the torsional stiffness are given by (Timoshenko and Goodier, 1951)

$$EI_1 = \frac{1}{12}Eht^3, \ EI_2 = \frac{1}{12}Eh^3t, \ GJ = \frac{Eht^3}{6(1+v)}\left\{1 - \frac{192}{\pi^5}\frac{t}{h}\sum_{n=1}^{\infty}\frac{1}{(2n-1)^5}\tanh\left[\frac{(2n-1)\pi h}{2t}\right]\right\}, \quad (3)$$

where $h$ is the cross-sectional height, $t$ is its thickness, and $v$ is the Poisson's ratio. By projecting Eq. (2) along the unit vectors $e_1$, $e_2$, and $e_3$, respectively, one obtains six equilibrium equations:

$$N_1' - N_2\kappa_3 + N_3\kappa_2 = 0, \ N_2' - N_3\kappa_1 + N_1\kappa_3 = 0, \ N_3' - N_1\kappa_2 + N_2\kappa_1 = 0,$$
$$M_1' + M_n' - M_2\kappa_3 + M_3\kappa_2 - N_2 = 0, \ M_2' + (M_1 + M_n)\kappa_3 - M_3\kappa_1 + N_1 = 0, \quad (4)$$
$$M_3' - (M_1 + M_n)\kappa_2 + M_2\kappa_1 = 0.$$

In parallel, a unit quaternion $\mathbf{q}=(q_0, q_1, q_2, q_3)$ with $q_0^2 + q_1^2 + q_2^2 + q_3^2 = 1$ is introduced to describe the orientation of the rod's cross-section during deformation (Healey and Mehta, 2005; Yu and Hanna, 2019), which also relates the local material frame to the global material frame, as

$$\begin{bmatrix} \mathbf{e}_1 \\ \mathbf{e}_2 \\ \mathbf{e}_3 \end{bmatrix} = 2\begin{bmatrix} q_0^2 + q_1^2 - \frac{1}{2} & q_1q_2 + q_0q_3 & q_1q_3 - q_0q_2 \\ q_1q_2 - q_0q_3 & q_0^2 + q_2^2 - \frac{1}{2} & q_2q_3 + q_0q_1 \\ q_1q_3 + q_0q_2 & q_2q_3 - q_0q_1 & q_0^2 + q_3^2 - \frac{1}{2} \end{bmatrix}\begin{bmatrix} \mathbf{E}_1 \\ \mathbf{E}_2 \\ \mathbf{E}_3 \end{bmatrix} = [\mathbf{Q}]\begin{bmatrix} \mathbf{E}_1 \\ \mathbf{E}_2 \\ \mathbf{E}_3 \end{bmatrix}. \quad (5)$$

Since $\mathbf{e}_3$ is the unit tangent vector to the centerline, we have $\mathbf{p}' = \mathbf{e}_3$. By using Eq. (5), the relationship between the position vector and the unit quaternion can be obtained as

$$p_1' = 2(q_1q_3 + q_0q_2), \ p_2' = 2(q_2q_3 - q_0q_1), \ p_3' = 2(q_0^2 + q_3^2) - 1. \quad (6)$$

Substituting Eq. (5) into Eq. (1) leads to $[\mathbf{e}_1', \mathbf{e}_2', \mathbf{e}_3']^T = [\mathbf{K}][\mathbf{Q}][\mathbf{E}_1, \mathbf{E}_2, \mathbf{E}_3]^T$. The derivative of Eq. (5) gives $[\mathbf{e}_1', \mathbf{e}_2', \mathbf{e}_3']^T = [\mathbf{Q}'][\mathbf{E}_1, \mathbf{E}_2, \mathbf{E}_3]^T$. Thus, $[\mathbf{K}][\mathbf{Q}] = [\mathbf{Q}']$, and equating the elements of matrices on both sides, one obtains

$$q_0' = (-q_1\kappa_1 - q_2\kappa_2 - q_3\kappa_3)/2, \ q_1' = (q_0\kappa_1 - q_3\kappa_2 + q_2\kappa_3)/2,$$
$$q_2' = (q_3\kappa_1 + q_0\kappa_2 - q_1\kappa_3)/2, \ q_3' = (-q_2\kappa_1 + q_1\kappa_2 + q_0\kappa_3)/2. \quad (7)$$

Eqs. (4), (6), and (7) provide 13 governing equations for the Kirchhoff rod, which produces a well-posed boundary value problem (BVP) when supplemented with appropriate boundary conditions. The rod model can not only evaluate the internal forces and moments during deformation but can also predict the deformation process of the rod under external load, which has been widely used to study the mechanical behavior of various slender elastic structures (Dias and Audoly, 2014; Kodio et al., 2020; Sano et al., 2022; Starostin and van der Heijden, 2022; Yu et al., 2023).

The curved-sided hexagrams studied in this paper are composed of curved edges and sharply rounded corners in practical fabrications. The initial curvatures at the joints of the edges and the corners are not continuous. In this regard, the curved-sided hexagrams are first divided into multiple segments, and then each segment is modeled as a Kirchhoff rod. Details on how these rings with different configurations are divided are shown in Fig. A1 in Appendix A. The rod models in different segments are coupled by imposing continuous or compatible boundary conditions, which form a multi-segment rod system (Sun et al., 2022). For a curved-sided hexagram divided into $m$ segments, there are $13m$ governing equations which contain $13m$ unknown variables. The unknown variables for the $j$-th segment of the curved-sided hexagram are denoted by $\left[N_i^{(j)}, \kappa_i^{(j)}, p_i^{(j)}, q_0^{(j)}, q_i^{(j)}\right]$, in which $i=1, 2, 3$, and $j=1, 2, \cdots, m$. To construct a unified multi-rod framework, we introduce the following dimensionless quantities,

$$\tilde{N}_i^{(j)} = \frac{N_i^{(j)} L^2}{GJ}, \ \tilde{M}_n = \frac{M_n L}{GJ}, \ \left(\tilde{\kappa}_i^{(j)}, \tilde{\kappa}_0^{(j)}\right) = \left(\frac{L\kappa_i^{(j)}}{2\pi}, \frac{L\kappa_0^{(j)}}{2\pi}\right),$$
$$\tilde{p}_i^{(j)} = \frac{p_i^{(j)}}{L}, \ \tilde{s}^{(j)} = \frac{s^{(j)}}{L\xi^{(j)}}, \ \frac{d(\bullet)}{ds^{(j)}} = \frac{1}{L\xi^{(j)}} \frac{d(\bullet)}{d\tilde{s}^{(j)}}, \tag{8}$$

where $L$ is the length of the curved edge, $\kappa_0^{(j)}$ is the initial curvature about $\mathbf{e}_1$ of the $j$-th segment, $\xi^{(j)} = l^{(j)} / L$ is a scaling factor to unify the normalized arc length $\tilde{s}^{(j)}$ of different segments into the same interval [0, 1] with $l^{(j)}$ being the length of the $j$-th segment. Note that the rounded corners of the rings carry the same bending moment $M_n$ in the natural state as that in the curved edges, such that the induced residual strain is continuous and uniform along the length direction. Then, with the definitions in Eq. (8), the dimensionless governing equations and boundary conditions for the multi-rod system can be obtained, as provided in Appendix A. The multi-rod system can be solved using various numerical continuation methods. Here, we use the Continuation Core and Toolboxes (COCO) (Dankowicz and Schilder, 2013) operated in MATLAB to handle the BVPs of the different multi-rod systems. Additionally, we employ a bending method, whose detail will be clarified later, to trigger the state transition of the ring with the bending angle Θ as the continuation parameter in the numerical implementation. The bending angle will vary from 0 to π, corresponding to the initial and final states, respectively. The associated equilibrated external bending moments applied at specific locations on the ring are also computed and presented. The accuracy of the numeral results is verified through independent FEA simulations with details presented in Appendix B.

## 3. Transition behavior of curved-sided hexagrams

To identify the transition rules between different equilibrium states of the curved-sided hexagram, in this section, we first study the transition behavior of each of the four equilibrium states using simulations based on the multi-segment Kirchhoff rod model and FEA by considering different natural curvatures and loading positions for the external bending stimuli. Each of the four equilibrium states, i.e., the star hexagram, the daisy hexagram, the 3-loop line, and the 3-loop "8", have an edge length of $L = 200$ mm, a cross-sectional height of $h = 2$ mm, and a thickness of $t = 0.5$ mm, corresponding to $h/t = 4$ in Part I. As noted, the curved-sided hexagrams considered in this part have rounded corners between adjacent curved edges to facilitate the numerical and experimental implementations. To reduce the effect of corner size, we employ a small corner radius $r = 1$ mm.

For each case, two loading positions of the external bending loads are considered, i.e., equal and opposite prescribed rotations at a pair of corners across the ring from each other or at the middle of a pair of edges opposite each other, which are referred to as "corner bend" and "edge bend", respectively. The magnitude of each externally imposed rotation is $\Theta$ and the associated moment is denoted by $M$ with the work done by the external stimuli as $W = 2\int_0^{\Theta} Md\Theta$. The elastic energy of the ring is represented by $U$, and its definition is provided in Appendix A.

*3.1. Natural curvature-dependent transitions from the star hexagram under corner bend*

We begin by studying the transition behavior of the star hexagram to other states for a range of natural curvatures under corner bend. The transition behavior is highly nonlinear and rich in phenomena. We will use the examples in this subsection to discuss the transition phenomena in some detail, starting with the case when the initial state of the star hexagram is stress-free (and necessarily stable).

**(a) Folding process of a star hexagram with $L\kappa_n/2\pi = 1/3$ (stress-free initial state)**

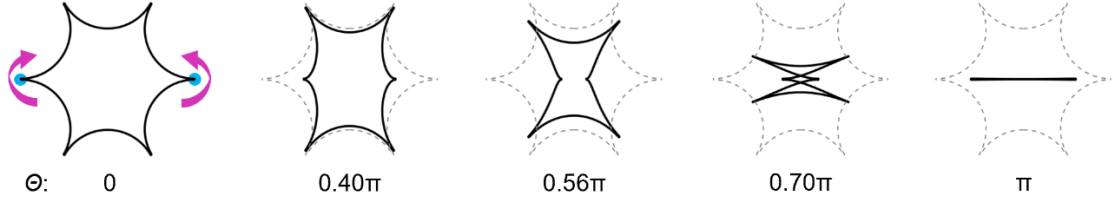

**(b) Inversion process of a star hexagram with $L\kappa_n/2\pi = 0.45$**

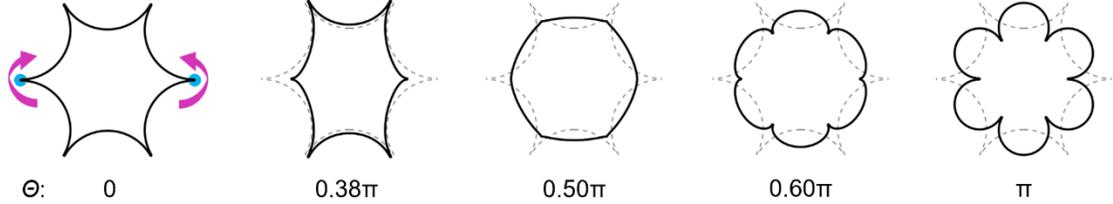

**(c) Star hexagram with natural curvature smaller than the initial curvature**

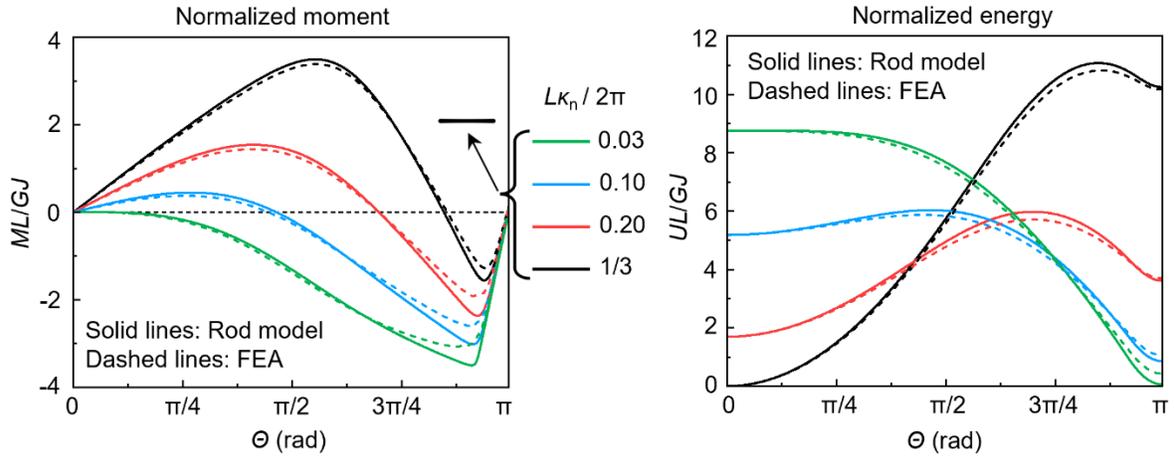

**(d) Star hexagram with natural curvature larger than the initial curvature**

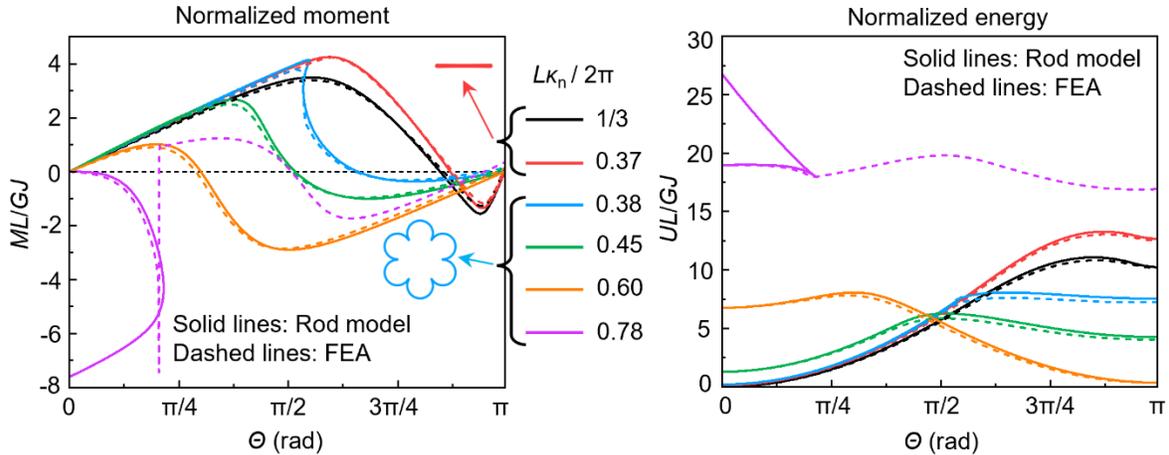

**Fig. 3.** Transition behavior of star hexagrams with different natural curvatures under corner bend. (a) Folding process (top view) of a star hexagram with a dimensionless natural curvature $L\kappa_n/2\pi = 1/3$, which has a stress-free initial state. (b) Inversion process (top view) of a star hexagram with a dimensionless natural curvature $L\kappa_n/2\pi = 0.45$. The blue dots denote the loading positions. (c) Normalized moment and energy versus bending angle of star hexagrams with natural curvatures smaller than the initial curvature. (d)

Normalized moment and energy versus bending angle of star hexagrams with natural curvatures larger than the initial curvature. The star hexagram with dimensionless natural curvature in the range (0.03, 0.37) folds into the 3-loop line, and in the range (0.38, 0.78) inverts into the daisy hexagram, as indicated by the insets. All the examples in this paper are for rods having a rectangular cross-section with $h/t = 4$ and $v = 1/3$.

The star hexagram has an edge radius of $3L/2\pi$ in its initial state, which corresponds to a dimensionless initial curvature $L\kappa_0/2\pi = 1/3$. As shown in Fig. 3(a), the star hexagram with a dimensionless natural curvature equal to its dimensionless initial curvature, $L\kappa_n/2\pi = 1/3$, which is stress-free in the initial state, transitions, or folds, to a 3-loop line when the bending angle reaches $\pi$. The isometric view of the folding process is provided in Fig. A3 in Appendix C. Examining the curves in Figs. 3(c) or 3(d) for this natural curvature, one sees that the transition solution has a monotonically increasing bending angle, $\Theta$, from 0 to $\pi$. This means that for loading conditions having prescribed increasing $\Theta$, the transition is stable with no dynamic snapping, assuming no bifurcation from this transition solution occurs (which is not considered in this paper).

Under prescribed increasing $\Theta$, the moment $M$ first increases, then decreases becoming negative, and then increases to zero in the final folded 3-loop line state. If the external stimuli were a pair of prescribed equal and opposite moments, $M$, then dynamic snapping would occur when the maximum moment is attained at $\Theta \cong 0.56\pi$. We have made no effort to study the snapping process under a prescribed value of the maximum $M$, and it is even possible that no stable static solution exists for that moment. Note that the external bending stimuli in our experiments have all been applied by hand, and in many cases, we have seen snapping in the transition process although no snapping is predicted for prescribed rotation in the numerical models. The discrepancy almost certainly stems from the fact that the human hand is not sufficiently rigid to apply a prescribed rotation.

There is an energy barrier between a stable star hexagram and any other state to which it might transition. For the case of the stress-free star hexagram (with $L\kappa_n/2\pi = 1/3$) just discussed, the energy barrier between it and the 3-loop line state is readily ascertained from the curve for this natural curvature in Fig. 3(c) or Fig. 3(d). The intermediate point in the transition solution where $M=0$, at $\Theta \cong 0.8\pi$, is an unstable static equilibrium state. If there were no external constraints on the deformed ring in this state, the ring would either snap back to the star hexagram state or snap to the 3-loop line state, depending on the disturbing perturbation. In other words, the ring could snap to the 3-loop line state with no further expenditure of external work on the ring. The energy barrier is the work $W$ required to bring the ring to the intermediate unstable equilibrium state, which equals the energy difference between the intermediate state and the initial state. For a star hexagram in the stress-free state, the elastic energy $U$ in Fig. 3(c) is equal to the work done by the external bending loads, and thus $U$ at $\Theta \cong 0.8\pi$ is the energy barrier. The point where $M = 0$ corresponds to the point where $dU/d\Theta = 0$ and where $U$ is a maximum in the second plot in Fig. 3(c). Thus, the energy barrier $\Delta U$ is $\Delta UL/GJ \cong 11$ for the star hexagram to 3-loop line transition when the initial state of the star hexagram is stress-free. This energy barrier should be regarded as an upper bound to the energy needed to transition from the star hexagram state to the 3-loop line state because there may exist other paths with other intermediate unstable static equilibrium states that can be activated by other stimuli requiring less work to reach the 3-loop line state.

Now consider the star hexagram in Fig. 3(b) which has a dimensionless natural curvature $L\kappa_n/2\pi = 0.45$. In the initial state, this star hexagram supports a uniform bending moment that increases the radius of the arc from its natural state, and as can be seen in Fig. 3(b), the star hexagram now undergoes an inversion transition into the daisy hexagram. The isometric view of the inversion process is provided in Fig. A3 in Appendix C. With reference to the curve in Fig. 3(d) relevant to this case, one sees again that the transition occurs monotonically with no snapping when the bending angle $\Theta$ is prescribed to increase monotonically from 0 to $\pi$. The intermediate unstable static equilibrium state occurs at $\Theta \cong 0.5\pi$, and the energy barrier between the star hexagram and daisy hexagram states is now the difference between $U$ at $\Theta \cong 0.5\pi$ and at $\Theta = 0$, resulting in $\Delta UL/GJ \cong 4$.

These two examples for the star hexagram illustrate that it can be made to transition to different states under the same external stimuli by rationally selecting the natural curvature. Later in this subsection, it will be demonstrated that the outcome of the transition can also be influenced by shifting the location of the loading points of the external stimuli. To give further insight into how the natural curvature influences transition, we have computed the transition behavior of the star hexagram under corner bend with a wider set of natural curvatures covering the full stability range of the star hexagram. The stability range for the dimensionless natural curvature of the star hexagram obtained in Part II is (0.03, 0.78), which is in excellent agreement with that determined in Part I, i.e., (0.025, 0.78), for $h/t = 4$ and $\nu = 1/3$. The outcomes of these additional calculations (all performed with both the rod model and FEA) are plotted in Figs. 3(c) and 3(d). In the range of dimensionless natural curvature, $L\kappa_n/2\pi$, given by (0.03, 0.37), the star hexagram transitions to the 3-loop line state, and the behavior is qualitatively similar to that described for the stress-free state, i.e., under prescribed monotonically increasing $\Theta$, there is no snapping, and the energy barrier is equal to $U_{max} - U(0)$. The maximum $U$ for the lower stability limit of the natural curvature range, $L\kappa_n/2\pi = 0.03$, occurs in the initial state at $\Theta = 0$, implying no energy barrier, as it must. Moreover, the associated $M$ variation with $\Theta$ appears to have a downward curvature at $\Theta = 0$, implying that the state at the lower limit, $L\kappa_n/2\pi = 0.03$, is unstable. In Part I, it was noted that the stability of the upper and lower limits of the range of stable behavior required further post-bifurcation analysis to ascertain their stability. The analysis of the transition in this part sheds light on the stability of these limiting states.

In the range of dimensionless natural curvature, (0.38, 0,78), in Fig. 3(d), the star hexagram transforms to the daisy hexagram. At the upper stability limit, $L\kappa_n/2\pi = 0.78$, the prescribed bending angle, $\Theta$, no longer monotonically increases under external stimuli. Moreover, the solution continuation algorithm for the rod model gives a different result for the transition state than that for the FEA simulation at the upper limit. The FEA predicts a transition involving snapping to the daisy hexagram state while the rod model under prescribed $\Theta$ predicts snapping at $\Theta \cong \pi/4$ without indicating to what state the rod would snap. The continuation method for the rod takes the solution back to a deformed star hexagram state with $\Theta$ decreasing back to zero. No effort has been made in this paper to resolve the uncertainty in the behavior at the upper limit of the stability range of the star hexagram—that remains for future study. However, both methods indicate that the $M$ vs. $\Theta$ relation has a negative curvature at $\Theta = 0$ implying that the star hexagram at the upper stability limit is also unstable.

One additional observation about the results in Figs. 3(c) and 3(d) deserves mentioning. With the exception of the limits of the range of dimensionless natural curvature, i.e., 0.03 and 0.78, the $U$ vs. $\Theta$ relation for the star hexagram state at $\Theta = 0$ has a positive second derivative, consistent with it being a stable state. At $\Theta = \pi$, the transition states, either the 3-loop line state or the daisy hexagram state depending on the natural curvature, one sees that the $U$ vs. $\Theta$ relation also has a positive second derivative, consistent with the transition states being stable in their respective ranges of natural curvature, as shown in Part I. It should also be reported that the numerical simulations with the rod and FEA models generating the plots in Fig. 3 and later plots in Part II have been conducted using increments of 0.01 in $L\kappa_n/2\pi$ over the relevant range of interest.

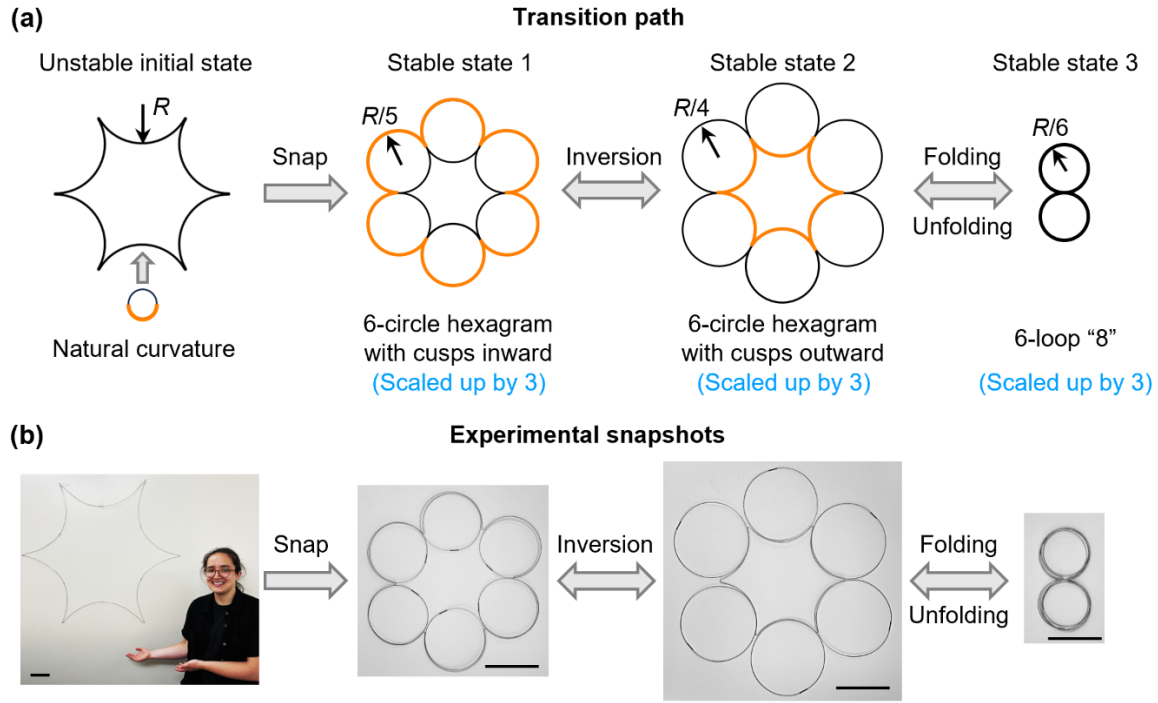

**Fig. 4.** Transitions between equilibrium states of an unstable star hexagram with a high natural curvature. (a) The unstable star hexagram (with edge radius $R=3L/2\pi$) snaps to a 6-circle hexagram with cusps pointing inward, which can further transform into the other two equilibrium states, i.e., a 6-circle hexagram with cusps pointing outward and a 6-loop "8". The natural curvature of one segment of the unstable star hexagram forms a roughly 1.5-loop circle with radius $2R/9$. The orange portions indicate where the rods overlap. (b) Experimental snapshots of the unstable star hexagram and its three equilibrium states. Note that the initial state (star hexagram) is unstable, but it is fixed at all corners to show the geometry. Once the constraints are removed, the unstable star hexagram spontaneously snaps to the 6-circle hexagram with cusps pointing inward. Scale bars: 10 cm.

Our experiments (and simulations) show that when the natural curvature is far above the upper stability limit for the star hexagram state (e.g., $L\kappa_n/2\pi = 1.5$), the initially unstable star hexagram (with edge radius $R=3L/2\pi$) snaps directly into a stable 6-circle hexagram configuration with two-layer overlapping outer edges, and their cusps pointing inward (i.e., stable state 1, in which the radius of the small circle is $R/5$), as shown in Fig. 4(a). This configuration can further invert into the other stable 6-circle hexagram with two-layer overlapping inner edges and cusps pointing outward (i.e., stable state 2, in which the radius of the small circle is $R/4$). The stability ranges for these two equilibrium states have been studied in Part I, and the results indicate that neither of them overlaps with the stability range of the star hexagram (if $h/t = 4$). Additionally, the two 6-circle hexagrams can both be folded into a stable 6-loop "8" (i.e., stable state 3, in which the radius of the small circle is $R/6$) if $L\kappa_n/2\pi = 1.5$. Experimental snapshots of the unstable star hexagram and its three equilibrium states are shown in Fig. 4(b), and the complete transition process between the three equilibrium states is presented in Video 2 in the Supplementary Materials. Experimental details on how to fabricate these 6-circle hexagrams are provided in Fig. S1 in the Supplementary Materials. Note that the unstable initial star hexagram state in the experiment is fixed at its corners. Once the constraints are removed, it spontaneously snaps to the 6-circle hexagram with cusps pointing inward.

*3.2. Natural curvature-dependent transitions from the star hexagram under edge bend*

Next, we study the transition behavior of star hexagrams under edge bend over the range of natural curvatures for which it is stable. As in the previous subsection, the stress-free star hexagram is first examined, as seen in Fig. 5(a). Interestingly, unlike the case of corner bend in which the stress-free star hexagram (with a dimensionless natural curvature $L\kappa_n/2\pi =1/3$) folds into a 3-loop line, under edge bend it transitions by inversion to the daisy hexagram state when the bending angle, $\Theta$, reaches $\pi$. Isometric views of the transition processes are provided in Fig. A4 in Appendix C. This demonstrates that the transition states of star hexagrams depend not only on the natural curvature but also on the loading positions of the external stimuli. The associated normalized moment and energy curve of the stress-free star hexagram under edge bend are plotted in Figs. 5(c) and 5(d). It is seen that the prescribed bending angle, $\Theta$, monotonically increases during the transition process, and the associated moment first increases, then decreases below zero, and finally increases back to zero as $\Theta$ varies from 0 to $\pi$. The energy barrier to be overcome to transition from the star hexagram to the daisy hexagram is $\Delta UL/GJ \cong 11$, essentially the same as for the transition to the 3-loop line state under corner bend.

**(a) Inversion process of a star hexagram with $L\kappa_n/2\pi = 1/3$ (stress-free initial state)**

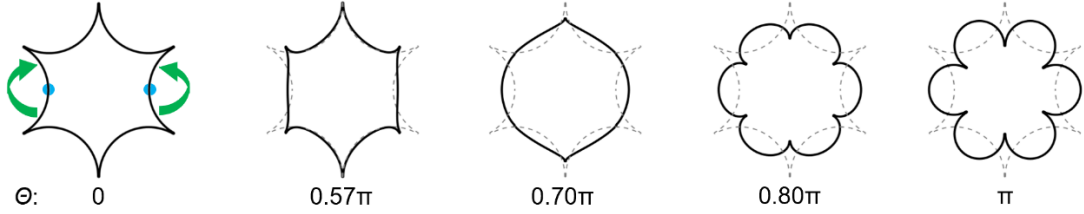

$\Theta$:   0     0.57π     0.70π     0.80π     π

**(b) Folding process of a star hexagram with $L\kappa_n/2\pi = 0.20$**

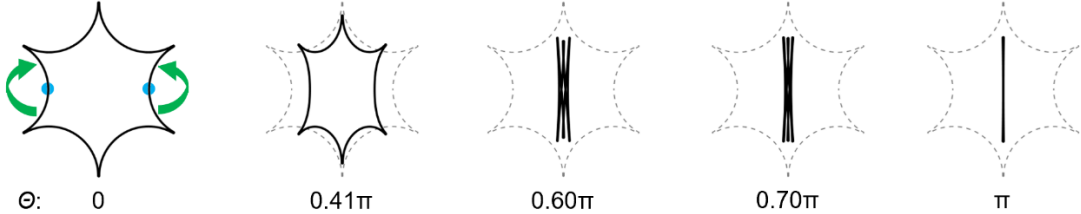

$\Theta$:   0     0.41π     0.60π     0.70π     π

**(c) Star hexagram with natural curvature smaller than the initial curvature**

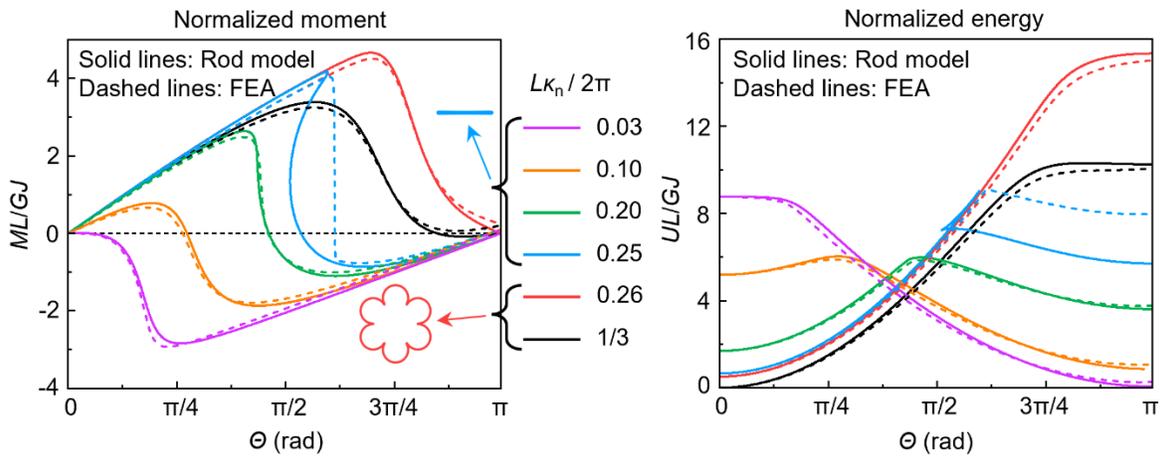

**(d) Star hexagram with natural curvature larger than the initial curvature**

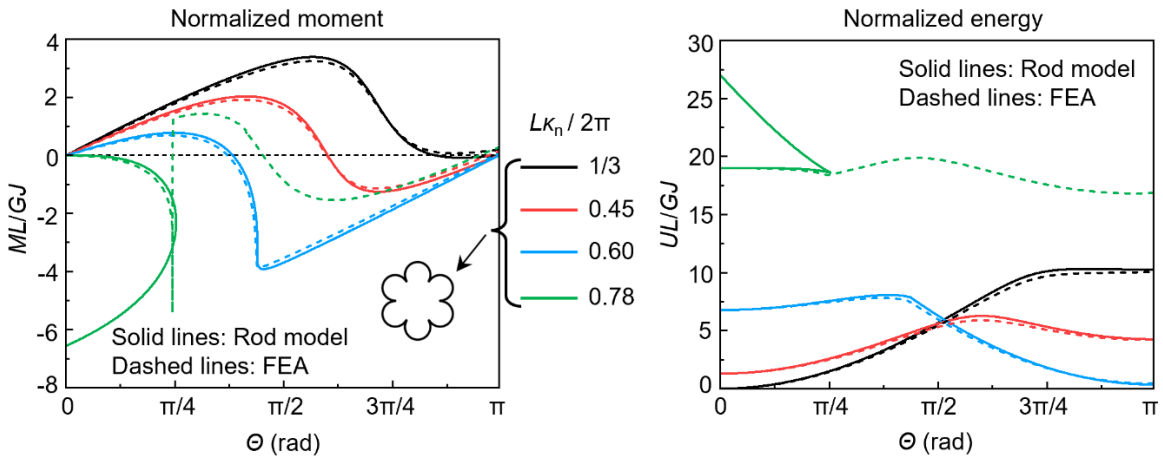

**Fig. 5.** Transition behavior of star hexagrams with different natural curvatures under edge bend. (a) Inversion process (top view) of a star hexagram with a dimensionless natural curvature $L\kappa_n/2\pi = 1/3$, which has a stress-free initial state. (b) Folding process (top view) of a star hexagram with a dimensionless natural curvature $L\kappa_n/2\pi = 0.20$. The blue dots denote the loading positions. (c) Normalized moment and energy versus bending angle of star hexagrams with natural curvatures smaller than the initial curvature. (d)

Normalized moment and energy versus bending angle of star hexagrams with natural curvatures larger than the initial curvature. The star hexagram with dimensionless natural curvature in the range (0.03, 0.25) folds into the 3-loop line, and in the range (0.26, 0.78) inverts into the daisy hexagram, as indicated by the insets.

Fig. 5(b) shows the transition process of the star hexagram with a dimensionless natural curvature $L\kappa_n/2\pi = 0.20$, under edge bend, which carries a uniform bending moment that reduces the radius of the arc from its natural state. The star hexagram folds into the 3-loop line state in this case, and the corresponding moment and energy curves during transition for this natural curvature are shown in Fig. 5(c). One can see that the prescribed bending angle increases monotonically from 0 to $\pi$ during the transition, and an intermediate unstable static equilibrium state occurs at $\Theta \cong 0.45\pi$ where $M = 0$. The energy barrier between the star hexagram state with $L\kappa_n/2\pi = 0.20$ and the 3-loop line state equals the energy difference between the intermediate state at $\Theta \cong 0.45\pi$ and the initial state $\Theta = 0$, resulting in $\Delta UL/GJ \cong 4$.

The transition behavior under edge bend stimuli of the star hexagram with other dimensionless natural curvatures within its stability range, (0.03, 0.78), is further examined in Figs. 5(c) and 5(d). In the dimensionless natural curvature range, (0.03, 0.25), the star hexagram folds into the 3-loop line configuration. In the upper end of this range, the prescribed bending angle, $\Theta$, does not increase monotonically during the transition, such as the case with $L\kappa_n/2\pi = 0.25$. Instead, $\Theta$ first increases, then decreases, and finally increases to $\pi$. For this case, the ring would snap dynamically at the point where $\Theta$ has a vertical slope ($\Theta \cong 0.62\pi$), as indicated by the dashed line curve generated by the FEA solution. For the star hexagram with dimensionless natural curvature within the range (0.26, 0.78), the transition is an inversion into a daisy hexagram. Except at the upper end of this range, the prescribed bending angle increases monotonically from 0 to $\pi$ during transition. At the lower and upper stability limits, i.e., $L\kappa_n/2\pi = 0.03$ and 0.78, the star hexagrams in the initial state are unstable to the edge bend stimuli, similar to the finding for corner bend.

*3.3. Brief summary of transitions of the four basic equilibrium states*

Transition calculations for the star hexagram, as reported above, have been performed for the daisy hexagram, the 3-loop line, and the 3-loop "8" for a range of natural curvatures under corner and edge bend. The results are presented in the Supplementary Materials in Figs. S2–S7. The outcomes of these simulations are summarized in Figs. 6 and 7, which will be reviewed briefly below. Videos showing numerical realizations of the transitions available are designated as Videos 3 and 4 in the Supplementary Materials.

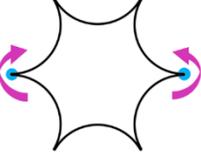

**Fig. 6.** Transition states of (a) the star hexagram and (b) the daisy hexagram with different natural curvatures within the stability range under corner bend and edge bend ($h/t = 4$ and $v = 1/3$). The blue dots denote the loading positions.

Transitions from the star hexagram for natural curvatures falling within, and at the limits, of the range in which it is stable have already been discussed, and these are summarized in Fig. 6(a). Within a small range of dimensionless curvature, including the initial curvature 1/3 for which the star hexagram state is stress-free, i.e., (0.26, 0.37), the star hexagram folds into the 3-loop line under corner bend but inverts into the daisy hexagram under edge bend. Like the star hexagram, the transition states from the daisy hexagram and from the 3-loop "8" depend on the natural curvature as well as the loading position. However, the transition state from the 3-loop line in Fig. 7(a) depends only on the natural curvature. Fig. 6(b) shows that the daisy hexagram with a dimensionless natural curvature in the range (0.32, 0.62) always inverts into the star hexagram, and in the range (0.70, 1.09) always folds into the 3-loop "8". When the dimensionless natural curvature of the daisy hexagram falls in the range (0.63, 0.69), it inverts to the star hexagram under corner bend, but folds into the 3-loop "8" under edge bend. Experimental demonstrations of the loading position-dependent transitions of a nearly stress-free star hexagram and a nearly stress-free daisy hexagram are provided in Video 5 in the Supplementary Materials. For the 3-loop line in Fig. 7(a), regardless of the bending locations, the 3-loop line inverts into a 3-loop "8" with edge contact in the middle in the range (0, 0.73), while it inverts into a 3-loop "8" without edge contact in the range (−0.73, 0). Note that for a stress-free 3-loop line ($L\kappa_n/2\pi = 0$), the edges of the obtained 3-loop "8" contact in the middle. As shown in Fig. 7(b), the 3-loop "8" inverts into a 3-loop line within the curvature range (0.64, 0.74), and folds into a 6-loop "8" within the curvature range (1.13, 1.38). When the dimensionless natural curvature is within the range (0.75, 1.12), the 3-loop "8" inverts to the 3-loop line under corner bend, but folds into a 6-loop "8" under edge bend. The results in Figs. 6 and 7 provide heuristic design and loading strategies for the curved-sided hexagram to achieve desired state transition behavior, as will be illustrated in the next section.

**(a) Transition states of the 3-loop line**

| Bending positions at the initial state | Dimensionless natural curvature ranges ||
|---|---|---|
| | (-0.73, 0) | (0, 0.73) |
| [corner bend diagram] | [diagram] | [diagram] |
| [edge bend diagram] | [diagram] | [diagram] |

**(b) Transition states of the 3-loop "8"**

| Bending positions at the initial state | Dimensionless natural curvature ranges |||
|---|---|---|---|
| | (0.64, 0.74) | (0.75, 1.12) | (1.13, 1.38) |
| [corner bend diagram] | [diagram] | [diagram] | [diagram] |
| [edge bend diagram] | [diagram] | [diagram] | [diagram] |

**Fig. 7.** Transition states of (a) the 3-loop line and (b) the 3-loop "8" with different natural curvatures within the stability range under corner bend and edge bend ($h/t = 4$ and $v = 1/3$). The blue dots denote the loading positions.

## 4. Transitions between four stable equilibrium states

Having revealed the transition behavior from each of the four basic equilibrium states when the natural curvature is in the range such that the basic state is stable, it is irresistible to consider an example when all four basic states are mutually stable and to analyze the transitions among all four states. Based on Figs. 6 and 7, we can obtain that all four basic states are stable when the dimensionless natural curvature, $L\kappa_n/2\pi$, falls within the range (0.64, 0.73), which is in great agreement with that predicted in Part I, namely (0.63, 0.75) for rods with a rectangular cross-section of $h/t = 4$ and $v = 1/3$. A stress-free daisy hexagram having dimensionless natural curvature, $L\kappa_n/2\pi = 2/3$, sits in this range, and we will use it as the starting configuration to demonstrate the transitions between all the four states.

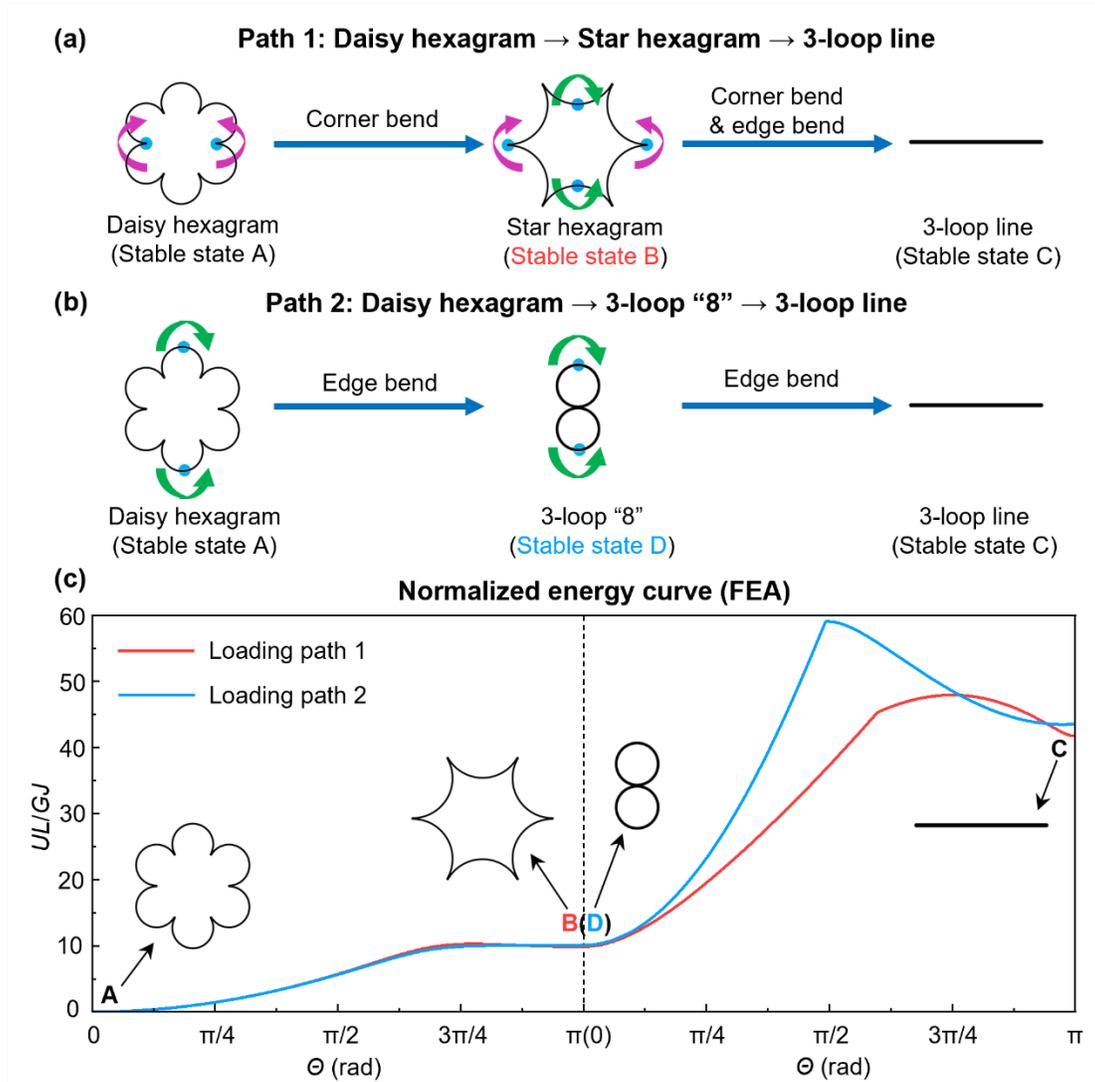

**Fig. 8.** Transition paths between the four stable equilibrium states of the curved-sided hexagram illustrated by starting with a stress-free daisy hexagram state ($L\kappa_n/2\pi = 2/3$). (a) Transition from the daisy hexagram to the star hexagram and then to the 3-loop line. (b) Transition from the daisy hexagram to the 3-loop "8" and then to the 3-loop line. (c) Energy landscapes of the transition paths.

In Fig. 6(b), it was shown that the daisy hexagram with a dimensionless natural curvature 2/3 can invert into the star hexagram under corner bend, and it can also fold into the 3-loop "8" under edge bend. Thus, in principle, we only need to explore how to obtain the 3-loop line configuration from either the star hexagram or the 3-loop "8". However, we will compute the transitions from the daisy hexagram state via each of the two intermediate states along the two paths depicted in Figs. 8(a) and 8(b). For path 1, we first invert the stress-free daisy hexagram into the star hexagram by bending at its corners, and then we apply bending loads at corners and edges simultaneously to fold the star hexagram into the 3-loop line. Note that if the bending loads had been applied only at a pair of corners or edges, the star hexagram would invert back to the daisy hexagram, as demonstrated in Fig. 6(a). For path 2, the 3-loop "8" obtained from the stress-free daisy hexagram can be inverted into the 3-loop line by directly applying bending loads at its edges. The energy landscapes of the two transition paths predicted by FEA are illustrated in Fig. 8(c). There is a local (or global for the daisy) minimum of $U$ at each of the basic states, as required by the fact that each of the states is stable. Moreover, one sees that the strain energy of the star hexagram state (stable state B) is identical to that of the 3-loop "8" state (stable state D), i.e., $U=4\pi^2 EI_1/(3L)$, but expressed in terms of the dimensionless $UL/GJ$. At the final stable state C, the strain energy of the 3-loop line obtained by the two paths should also be the same. Note that the minor strain energy discrepancy at stable state C is due to the slightly different 3-loop line configurations achieved through different loading paths potentially caused by numerical errors (such as damping and meshing). These transition paths, as well as those in the earlier section, can be reversed by selecting appropriate bending locations, thus forming a closed-loop transition sequence as shown in Fig. 1. Reversible behavior is also demonstrated experimentally in Video 1 in the Supplementary Materials. Finally, it is important to mention that contact algorithms have not been needed in any of the simulations in this paper. The transitions discussed in this paper have not involved any instances where the motion of the rods causes them to 'pass through each other', although this might seem surprising given the complexity of some of the transition histories.

## 5. Concluding remarks

This paper has presented a study of the transitions between the four basic equilibrium states, i.e., the star hexagram, the daisy hexagram, the 3-loop line, and the 3-loop "8", of a curved-sided hexagram. Simulations have been based on a multi-segment Kirchhoff rod model and finite element analysis, and experiments validating nearly every aspect studied theoretically have been performed. The rod model was solved using a numerical continuation method to transition from one state to another, and it has been validated by independently performed finite element simulations. The effects of the natural curvature of the rods used to form the curved-sided hexagram and the positions of the external bending stimuli on the transition behavior from each of the four basic equilibrium states have been determined for natural curvatures falling with the stability range of each of the states. The results show that both natural curvature and the location of the external stimuli are important in determining the transition response. For example, the star hexagram with a dimensionless natural curvature within the range (0.26, 0.37) can fold into the 3-loop line when bent at corners but invert into the daisy hexagram when bent at edges. By contrast, when the natural curvature is outside this range (but still within the stability range), the star hexagram will either fold into the 3-loop line or invert into the daisy hexagram, regardless of the bending positions. Similar phenomena are observed in the transitions of the daisy hexagram and the 3-loop "8". These findings provide guidelines for achieving the transitions between the four equilibrium states of the curved-sided hexagram, such as the loading strategies proposed in Section 4 for a curved-sided hexagram designed such that all four states are mutually stable.

The nonlinear phenomena exhibited by the curved-sided hexagram are unusually rich. For example, it has been shown that a star hexagram with a modestly large natural curvature is unstable and will snap into a stable 6-circle hexagram. This same star hexagram could be folded into an unstable 3-loop line state that could be stabilized if properly constrained. It is anticipated, but not established, that once the constraint on the 3-loop line is released it would snap to the six-circle hexagram, providing its own energy for the transition. Such compact and energetic packaging might have functional uses. Hopefully, the curved-sided hexagram with multiple equilibrium states studied in Parts I and II will add to the prospective for the design of multi-functional deployable and foldable structures.

## Acknowledgments

R.R.Z., L.L, J.D., S.L. acknowledge National Science Foundation Award CPS-2201344 and National Science Foundation Career Award CMMI-2145601 for the support of this work.

## Supplementary materials

Supplementary material associated with this article can be found in the online version.

# Appendix A. Governing equations and boundary conditions for the multi-rod system

The nondimensional governing equations for the $j$-th ($j=1, 2, \cdots, m$) segment of the multi-rod system are given by

$$\frac{d\tilde{N}_1^{(j)}}{d\tilde{s}^{(j)}} = 2\pi\xi^{(j)}\left(\tilde{N}_2^{(j)}\tilde{\kappa}_3^{(j)} - \tilde{N}_3^{(j)}\tilde{\kappa}_2^{(j)}\right), \quad \frac{d\tilde{N}_2^{(j)}}{d\tilde{s}^{(j)}} = 2\pi\xi^{(j)}\left(\tilde{N}_3^{(j)}\tilde{\kappa}_1^{(j)} - \tilde{N}_1^{(j)}\tilde{\kappa}_3^{(j)}\right),$$

$$\frac{d\tilde{N}_3^{(j)}}{d\tilde{s}^{(j)}} = 2\pi\xi^{(j)}\left(\tilde{N}_1^{(j)}\tilde{\kappa}_2^{(j)} - \tilde{N}_2^{(j)}\tilde{\kappa}_1^{(j)}\right),$$

$$\frac{d\tilde{\kappa}_1^{(j)}}{d\tilde{s}^{(j)}} = \frac{d\tilde{\kappa}_0^{(j)}}{d\tilde{s}^{(j)}} + \left[2\pi(\beta-1)\tilde{\kappa}_2^{(j)}\tilde{\kappa}_3^{(j)} + \frac{\tilde{N}_2^{(j)}}{2\pi}\right]\frac{\xi^{(j)}}{\alpha}$$

$$\frac{d\tilde{\kappa}_2^{(j)}}{d\tilde{s}^{(j)}} = \left\{2\pi\tilde{\kappa}_1^{(j)}\tilde{\kappa}_3^{(j)} - \left[2\pi\alpha(\tilde{\kappa}_1^{(j)} - \tilde{\kappa}_0^{(j)}) + \tilde{M}_n\right]\tilde{\kappa}_3^{(j)} - \frac{\tilde{N}_1^{(j)}}{2\pi}\right\}\frac{\xi^{(j)}}{\beta},$$

$$\frac{d\tilde{\kappa}_3^{(j)}}{d\tilde{s}^{(j)}} = \left\{\left[2\pi\alpha(\tilde{\kappa}_1^{(j)} - \tilde{\kappa}_0^{(j)}) + \tilde{M}_n\right]\tilde{\kappa}_2^{(j)} - 2\pi\beta\tilde{\kappa}_1^{(j)}\tilde{\kappa}_2^{(j)}\right\}\xi^{(j)},$$

$$\frac{d\tilde{p}_1^{(j)}}{d\tilde{s}^{(j)}} = 2\xi^{(j)}\left(q_1^{(j)}q_3^{(j)} + q_0^{(j)}q_2^{(j)}\right), \quad \frac{d\tilde{p}_2^{(j)}}{d\tilde{s}^{(j)}} = 2\xi^{(j)}\left(q_2^{(j)}q_3^{(j)} - q_0^{(j)}q_1^{(j)}\right),$$

$$\frac{d\tilde{p}_3^{(j)}}{d\tilde{s}^{(j)}} = 2\xi^{(j)}\left[(q_0^{(j)})^2 + (q_3^{(j)})^2 - \frac{1}{2}\right],$$

$$\frac{d\tilde{q}_0^{(j)}}{d\tilde{s}^{(j)}} = \pi\xi^{(j)}\left(-q_1^{(j)}\tilde{\kappa}_1^{(j)} - q_2^{(j)}\tilde{\kappa}_2^{(j)} - q_3^{(j)}\tilde{\kappa}_3^{(j)}\right), \quad \frac{d\tilde{q}_1^{(j)}}{d\tilde{s}^{(j)}} = \pi\xi^{(j)}\left(q_0^{(j)}\tilde{\kappa}_1^{(j)} - q_3^{(j)}\tilde{\kappa}_2^{(j)} + q_2^{(j)}\tilde{\kappa}_3^{(j)}\right),$$

$$\frac{d\tilde{q}_2^{(j)}}{d\tilde{s}^{(j)}} = \pi\xi^{(j)}\left(q_3^{(j)}\tilde{\kappa}_1^{(j)} + q_0^{(j)}\tilde{\kappa}_2^{(j)} - q_1^{(j)}\tilde{\kappa}_3^{(j)}\right), \quad \frac{d\tilde{q}_3^{(j)}}{d\tilde{s}^{(j)}} = \pi\xi^{(j)}\left(-q_2^{(j)}\tilde{\kappa}_1^{(j)} + q_1^{(j)}\tilde{\kappa}_2^{(j)} + q_0^{(j)}\tilde{\kappa}_3^{(j)}\right),$$

(A.1)

in which

$$\alpha = \frac{EI_1}{GJ}, \quad \beta = \frac{EI_2}{GJ}. \tag{A.2}$$

Schematics of the four equilibrium states of the curved-sided hexagram under bending at corners or at edges are shown in Figs. A1(a) and A1(b), respectively. It is seen that all the four states are axisymmetric, thus we can take one quarter of them (denoted by orange lines) in the analysis for simplicity. The quarter rings are further divided into multiple segments to ensure that the initial curvature of each segment is continuous, and thus each segment can be modeled as a Kirchhoff rod. For the star and daisy hexagrams, they are divided into four segments. For the 3-loop line and 3-loop "8", they are divided into two segments. An *m*-segment rod system has $13m$ governing equations, which require $13m$ boundary conditions to produce a well-posed BVP. At each boundary, there are two types of boundary conditions, i.e., translational boundary conditions and rotational boundary conditions. For rotational boundary conditions, it is more straightforward to define them using Euler angles. Therefore, we first define the rotational boundary conditions using Euler angle $(\varphi, \theta, \phi)$, and then convert them into quaternions. The relationships between the Euler angle follow the 3-2-3 rotation convention and the quaternions are given by (Healey and Mehta, 2005; Yu and Hanna, 2019)

$$q_0 = \cos\frac{\theta}{2}\cos\frac{\phi+\varphi}{2}, \quad q_1 = \sin\frac{\theta}{2}\sin\frac{\phi-\varphi}{2}, \quad q_2 = \sin\frac{\theta}{2}\cos\frac{\phi-\varphi}{2}, \quad q_3 = \cos\frac{\theta}{2}\sin\frac{\phi+\varphi}{2}. \quad (A.3)$$

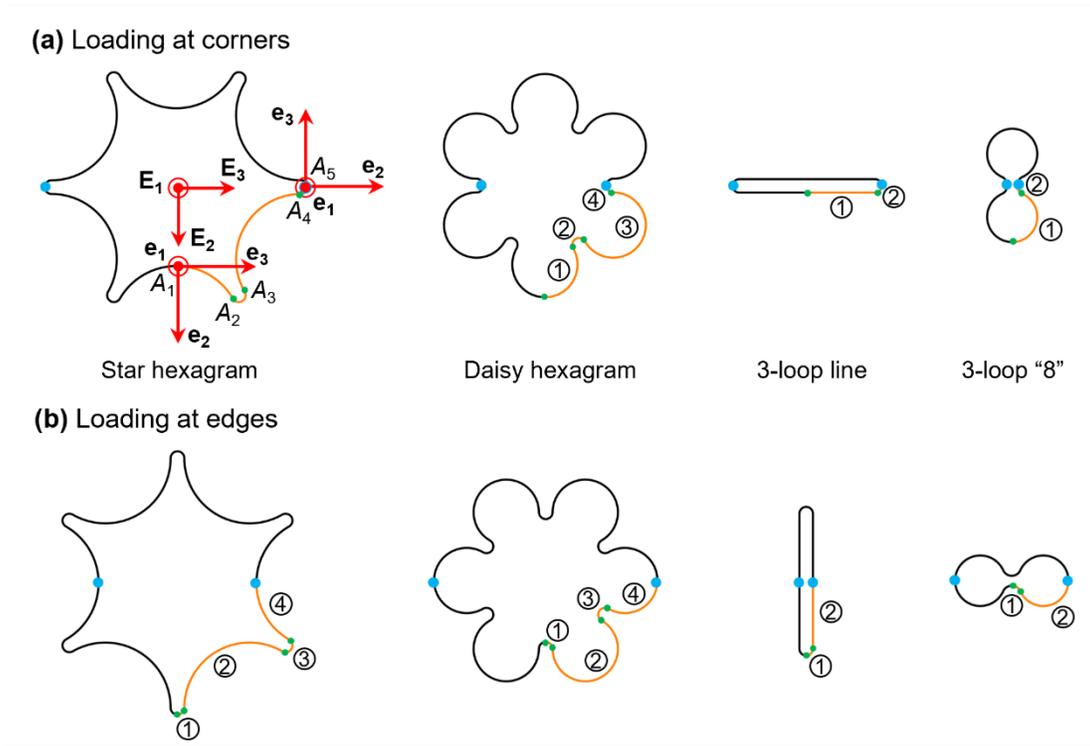

**Fig. A1.** Schematics of the four equilibrium states of the curved-sided hexagrams under bending (a) at corners and (b) at edges. The blue dots denote the loading positions, and the green dots represent the joints of adjacent segments. The circled numbers are the segment numbers. The local material frame ($e_1$, $e_2$, $e_3$) is attached to the centerline, and the global material frame ($E_1$, $E_2$, $E_3$) is located at the center of the rings.

Next, we take the star hexagram as an example to derive the boundary conditions for the multi-rod system under external bending loads. When the star hexagram is under external bending loads at the two corners, as shown in Fig. A1(a), it is free to rotate about $e_3$ at the left boundary of the first segment (i.e., point $A_1$). If we denote the unknown rotational angle as $\gamma$, the Euler angles are given by $\varphi = \gamma$, $\theta = 0$, $\phi = 0$. Thus, we have $q_0^{(1)} = \cos(\gamma/2)$, $q_1^{(1)} = 0$, $q_2^{(1)} = 0$, $q_3^{(1)} = \sin(\gamma/2)$. Since the rotational angle is not prescribed in this boundary, we have $M_3^{(1)} = 0$, leading to $\kappa_3^{(1)} = 0$. Thus, the rotational boundary conditions at the left boundary of the first segment are written as

$$q_1^{(1)}(0) = 0, \quad q_2^{(1)}(0) = 0, \quad \tilde{\kappa}_3^{(1)}(0) = 0. \tag{A.4}$$

Moreover, the left end of the first segment remains in the $\mathbf{e}_1$-$\mathbf{e}_3$ plane and is free to translate along $\mathbf{e}_2$. Therefore, the translational boundary conditions at the left boundary of the first segment can be expressed as

$$\tilde{p}_1^{(1)}(0) = 0, \quad \tilde{p}_3^{(1)}(0) = 0, \quad \tilde{N}_2^{(1)}(0) = 0. \tag{A.5}$$

At the right boundary of the fourth segment (i.e., point $A_5$) where the bending load is applied, the Euler angles are written as $\varphi = -\pi/2$, $\theta = \pi/2$, $\phi = \pi/2 - \Theta$, in which $\Theta$ is the prescribed bending angle. By using Eq. (A.3), the rotational boundary conditions at the right boundary of the fourth segment can be obtained as

$$q_0^{(4)}(1) = \frac{\sqrt{2}}{2}\cos\left(\frac{\Theta}{2}\right), \quad q_1^{(4)}(1) = \frac{\sqrt{2}}{2}\cos\left(\frac{\Theta}{2}\right),$$
$$q_2^{(4)}(1) = \frac{\sqrt{2}}{2}\sin\left(\frac{\Theta}{2}\right), \quad q_3^{(4)}(1) = -\frac{\sqrt{2}}{2}\sin\left(\frac{\Theta}{2}\right). \tag{A.6}$$

Additionally, under bending loads, the right end of the fourth segment has no displacement along $\mathbf{e}_3$ but is free to translate along $\mathbf{e}_1$ and $\mathbf{e}_2$. Note that the position vector $\mathbf{p}$ is defined on the global material frame. Therefore, the translational boundary conditions at the right boundary of the fourth segment are written as

$$\tilde{N}_1^{(4)}(1) = 0, \quad \tilde{N}_2^{(4)}(1) = 0, \quad \tilde{p}_2^{(4)}(1) = 0. \tag{A.7}$$

Eqs. (A.4)–(A.7) provide 13 boundary conditions for the governing equations of the star hexagram, which also apply for the other three equilibrium states, i.e., the daisy hexagram, 1-loop line, and 1-loop "8". The remaining $13(m-1)$ boundary conditions can be obtained by considering the force equilibrium and geometric compatibility at the joint of adjacent segments. For example, the boundary conditions at the joint $A_2$ of the star hexagram can be written as

$$\tilde{N}_1^{(1)}(1) = \tilde{N}_1^{(2)}(0), \quad \tilde{N}_2^{(1)}(1) = \tilde{N}_2^{(2)}(0), \quad \tilde{N}_3^{(1)}(1) = \tilde{N}_3^{(2)}(0),$$
$$\tilde{\kappa}_1^{(1)}(1) + \tilde{\kappa}_0^{(2)} = \tilde{\kappa}_1^{(2)}(0) + \tilde{\kappa}_0^{(1)}, \quad \tilde{\kappa}_2^{(1)}(1) = \tilde{\kappa}_2^{(2)}(0), \quad \tilde{\kappa}_3^{(1)}(1) = \tilde{\kappa}_3^{(2)}(0),$$
$$\tilde{p}_1^{(1)}(1) = \tilde{p}_1^{(2)}(0), \quad \tilde{p}_2^{(1)}(1) = \tilde{p}_2^{(2)}(0), \quad \tilde{p}_3^{(1)}(1) = \tilde{p}_3^{(2)}(0),$$
$$q_0^{(1)}(1) = q_0^{(2)}(0), \quad q_1^{(1)}(1) = q_1^{(2)}(0), \quad q_2^{(1)}(1) = q_2^{(2)}(0), \quad q_3^{(1)}(1) = q_3^{(2)}(0).$$
(A.8)

In a similar way, the remaining boundary conditions at the other joints can be obtained. Boundary conditions for the case of bending at edges are identical to the case of bending at corners, except for the initial curvature of each segment $\tilde{\kappa}_0^{(j)}$ in Eq. (A.8).

Further, the total elastic strain energy of the multi-rod system can be calculated as

$$U = 4\sum_{j=1}^{m} \left[ \int_s \frac{(M_1^{(j)} + M_n)^2}{2EI_1}ds + \int_s \frac{(M_2^{(j)})^2}{2EI_2}ds + \int_s \frac{(M_3^{(j)})^2}{2GJ}ds \right]$$
$$= 2\sum_{j=1}^{m} \left[ \begin{array}{l} \int_s EI_1(\kappa_1^{(j)} - \kappa_0^{(j)})^2 ds + \int_s EI_2(\kappa_2^{(j)})^2 ds + \int_s GJ(\kappa_3^{(j)})^2 ds \\ + \int_s 2(\kappa_1^{(j)} - \kappa_0^{(j)})M_n ds + \int_s \frac{M_n^2}{EI_1}ds \end{array} \right].$$
(A.9)

## Appendix B. Finite element simulations

When the natural curvature of the curved-sided hexagram is not equal to its initial curvature, the hexagram ring carries a uniform bending moment in the initial state, and the bending moment tends to bend the ring inwards or outwards. To simulate the effect of the initial bending moment on the transition behavior of the curved-sided hexagram, we consider a bilayer ring with opposite thermal expansion in the two layers, as shown in Fig. A2(a). Specifically, the inner layer has a negative coefficient of thermal expansion (CTE), $-\alpha_T$, and the outer layer has a positive CTE, $\alpha_T$, and the two coefficients are equal in magnitude. Thereby, when the bilayer undergoes a temperature change, $\Delta T$, the inner layer is under thermal contraction, and the outer layer is under thermal expansion. The thermal strain mismatch between the two layers produces a bending moment, which acts as the natural curvature induced bending moment. Thermal stress distribution across the thickness of the bilayer is shown in Fig. A2(b), and its resultant bending moment can be obtained as

$$M_T = \int_{-h/2}^{h/2}\int_{-t/2}^{0} \sigma_T x dx dz + \int_{-h/2}^{h/2}\int_{-t/2}^{0} (-\sigma_T)x dx dz = -\frac{1}{4}\sigma_T h t^2 = -\frac{1}{4}E\alpha_T \Delta T h t^2.$$
(B.1)

As noted in Section 2, the uniform bending moment carried in the curved-sided hexagram in the initial state is given by $M_n = EI_1(\kappa_0 - \kappa_n)$. By setting the two bending moments equal, the temperature change that needs to be prescribed to simulate the initial bending moment in the ring can be obtained as

$$\Delta T = -\frac{t(\kappa_0 - \kappa_n)}{3\alpha_T}. \qquad (B.2)$$

Note that for the star hexagram, the curvatures considered in Figs. 3 and 5 should take negative values to produce the corresponding temperature change using Eq. (B.2), because the reference curving direction of a positive curvature in star hexagram is defined to be opposite to that in daisy hexagram and 3-loop 8 (as shown in Fig. 1 in Part I). In addition, although such a bilayer can reproduce the initial bending moment in the ring, the initial elastic energy introduced by the thermal stress in the bilayer is not equal to that induced by natural curvature. Therefore, the elastic energy of the curved-sided hexagrams in FEA simulations are calculated as the sum of the external work, $W$, and the initial energy, $U_0$, induced by the natural curvature. Particularly, $U_0 = M_n^2 L_{total} / 2EI_1$, in which $L_{total}$ represents the total length of the curved-sided hexagram.

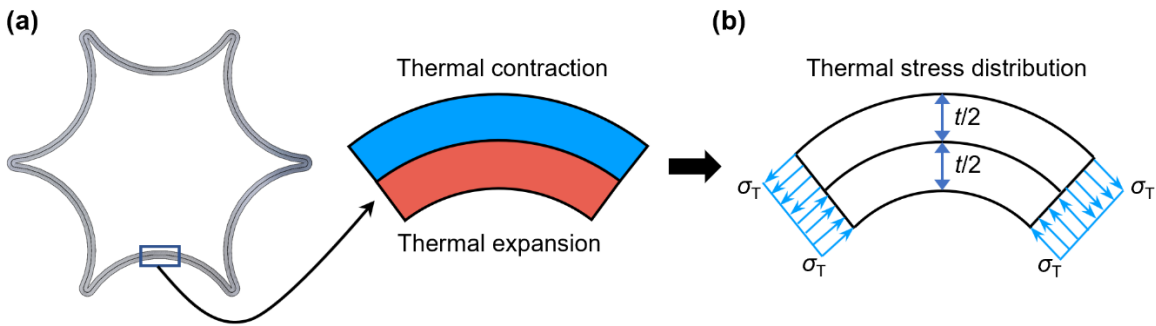

**Fig. A2.** Finite element implementation for the curved-sided hexagram. (a) Schematic of the bilayer ring considered in FEA whose inner layer is under thermal contraction, but the outer layer is under thermal expansion. (b) Thermal stress distribution across the thickness direction in the bilayer ring.

After obtaining the required temperature change to simulate the initial bending moment, transition behavior of curved-sided hexagrams under bending loads is simulated in the commercial software ABAQUS 2021 (Dassault Systèmes, France). The ring sizes are identical to those used in the rod model. C3D8R element is used in all simulations, and the mesh size is set as 0.3 mm. The Young's modulus, the coefficient of thermal expansion, and the Poisson's ratio are taken as $E$ = 200 GPa, $\alpha_T$ = 0.002 K$^{-1}$, and $v$ = 1/3, respectively. To trigger the transitions of the rings, a pair of bending angles are applied to the center of a set of opposite edges or a set of opposite corners. Further, a small damping factor $10^{-8}$ is used to stabilize the simulation.

## Appendix C. Isometric views of transition processes

Figs. A3 and A4 show the isometric views of transition processes of the star hexagram with different natural curvatures under corner bend and edge bend, respectively. It is clearly seen that the star hexagram always requires out-of-plane deformation to transition from one stable state to another, regardless of the loading method.

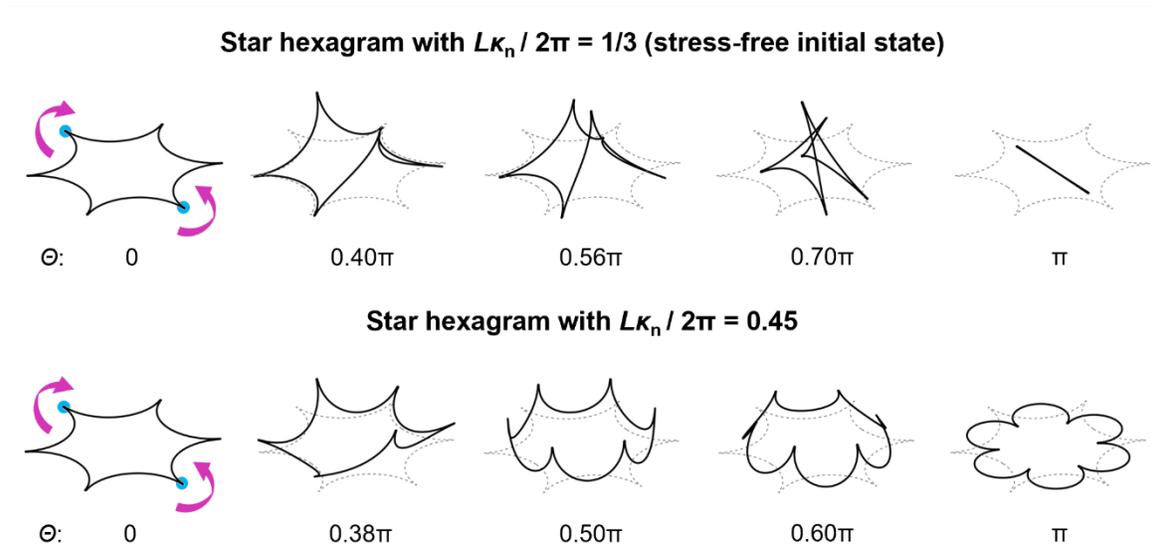

**Fig. A3.** Isometric views of transition processes of star hexagrams with different natural curvatures under corner bend. $L\kappa_n/2\pi$ is the dimensionless natural curvature, $\Theta$ is the bending angle, and the blue dots denote the loading positions.

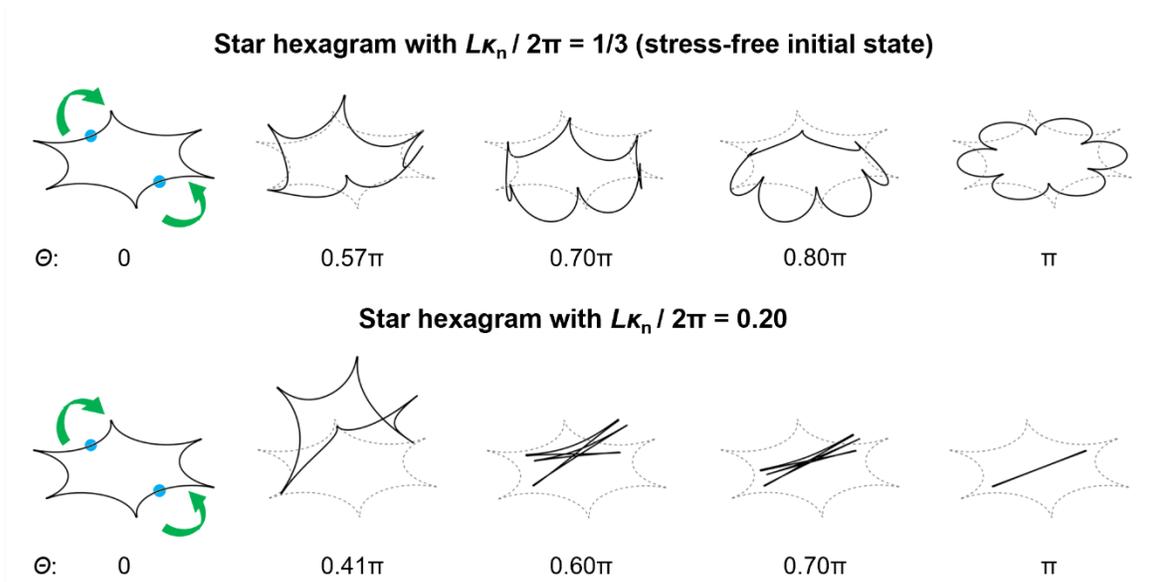

**Fig. A4.** Isometric views of transition processes of star hexagrams with different natural curvatures under edge bend. $L\kappa_n/2\pi$ is the dimensionless natural curvature, $\Theta$ is the bending angle, and the blue dots denote the loading positions.

Fu, H., Nan, K., Bai, W., Huang, W., Bai, K., Lu, L., Zhou, C., Liu, Y., Liu, F., Wang, J., Han, M., Yan, Z., Luan, H., Zhang, Y., Zhang, Y., Zhao, J., Cheng, X., Li, M., Lee, J.W., Liu, Y., Fang, D., Li, X., Huang, Y., Zhang, Y., Rogers, J.A., 2018. Morphable 3D mesostructures and microelectronic devices by multistable buckling mechanics. Nature Materials 17, 268-276.

Healey, T.J., Mehta, P., 2005. Straightforward computation of spatial equilibria of geometrically exact cosserat rods. International Journal of Bifurcation and Chaos 15, 949-965.

Jawed, M.K., Dieleman, P., Audoly, B., Reis, P.M., 2015. Untangling the mechanics and topology in the frictional response of long overhand elastic knots. Physical Review Letters 115, 118302.

Kodio, O., Goriely, A., Vella, D., 2020. Dynamic buckling of an inextensible elastic ring: Linear and nonlinear analyses. Physical Review E 101, 053002.

Leanza, S., Wu, S., Dai, J., Zhao, R.R., 2022. Hexagonal Ring Origami Assemblies: Foldable Functional Structures with Extreme Packing. Journal of Applied Mechanics 89, 081003.

Leanza, S., Zhao, R.R., Hutchinson, J.W., 2023. On the elastic stability of folded rings in circular and straight states. European Journal of Mechanics-A/Solids, 105041.

Li, Y., Pellegrino, S., 2020. A theory for the design of multi-stable morphing structures. Journal of the Mechanics and Physics of Solids 136, 103772.

Liu, M., Domino, L., de Dinechin, I.D., Taffetani, M., Vella, D., 2023a. Snap-induced morphing: From a single bistable shell to the origin of shape bifurcation in interacting shells. Journal of the Mechanics and Physics of Solids 170, 105116.

Liu, Z., Fang, H., Xu, J., Wang, K., 2023b. Digitized design and mechanical property reprogrammability of multistable origami metamaterials. Journal of the Mechanics and Physics of Solids 173, 105237.

Lu, L., Leanza, S., Dai, J., Sun, X., Zhao, R.R., 2023a. Easy snap-folding of hexagonal ring origami by geometric modifications. Journal of the Mechanics and Physics of Solids 171, 105142.

Lu, L., Leanza, S., Zhao, R.R., 2023b. Origami with Rotational Symmetry: A Review on Their Mechanics and Design. Applied Mechanics Reviews 75, 050801.

Manning, R.S., Hoffman, K.A., 2001. Stability of n-covered circles for elastic rods with constant planar intrinsic curvature. Journal of Elasticity 62, 1-23.

Melancon, D., Gorissen, B., García-Mora, C.J., Hoberman, C., Bertoldi, K., 2021. Multistable inflatable origami structures at the metre scale. Nature 592, 545-550.

Meng, Z., Liu, M., Yan, H., Genin, G.M., Chen, C.Q., 2022. Deployable mechanical metamaterials with multistep programmable transformation. Science Advances 8, eabn5460.

Miller, J.T., Lazarus, A., Audoly, B., Reis, P.M., 2014. Shapes of a suspended curly hair. Physical Review Letters 112, 068103.